# Analyzing Force Concept Inventory with Item Response Theory

Jing Wang

*Department of Physics and Astronomy, Eastern Kentucky University, Richmond, Kentucky 40475*

*Department of Physics, The Ohio State University, Columbus, Ohio 43210*

Lei Bao[a]

*Department of Physics, The Ohio State University, Columbus, Ohio 43210*

**Abstract**

Item Response Theory (IRT) is a popular assessment method used in education measurement, which builds on an assumption of a probability framework connecting students' innate ability and their actual performances on test items. The model transforms students' raw test scores through a nonlinear regression process into a scaled proficiency rating, which can be used to compare results obtained with different test questions. IRT also provides a theoretical approach to address ceiling effect and guessing. We applied IRT to analyze the Force Concept Inventory (FCI). The data was collected from 2802 students taking intro level mechanics courses at The Ohio State University. The data was analyzed with a 3-parameter item response model for multiple choice questions. We describe the procedures of the analysis and discuss the results and the interpretations. The analysis outcomes are compiled to provide a detailed IRT measurement metric of the FCI, which can be easily referenced and used by teachers and researchers for a range of assessment applications.

## I. Introduction

Assessment is an important component of education; it provides guidance to the learning process, measures achievement of the learners, and evaluates the effectiveness of teaching and learning activities.[1-3] To serve these goals, formal measurement models are created to provide explicit rules that integrate pieces of information drawn from assessment tasks into statistical structures for meaningful interpretation of data. Item response theory (IRT) is one of these modeling methods based on a statistical framework.[4] IRT attempts to provide examinee-independent question features such as question difficulty and question discrimination.[5] In general, testing is an interactive procedure between examinees and test items, and therefore, the measurement outcomes are affected by features of both the test takers and the test itself. By making certain assumptions of the examinee population, one can build models (referred as item response models) to describe and evaluate the characteristics of the questions. On the other hand, the tested models of the item features can be used to evaluate characteristics of different examinee groups that belong to the same population.

In this paper, we apply IRT methods to analyze test data of Force Concept Inventory (FCI)[6]. Based on students' overall performance on the FCI test, for each question we fit an item response model that describes the population's performance on the question and estimates the information on both the ceiling effect and the guessing chances. The IRT fits of FCI questions are compiled to provide a measurement metric for each of the FCI items, with which teachers and researchers can obtain more detailed information from their assessment data.

In the following sections, we will first review the basic framework of item response modeling and the model used in our study. We then discuss the design and data used in our study and evaluate the validity of the item response model used in the analysis. We summarize all the analysis results to form a measurement metric for FCI in graphs and tables, which can be conveniently used by teachers and researchers to compare with their assessment results. We will discuss examples on how to use the FCI-metric and provide suggestions for physics teachers. The two sets of terms, "item" and "question", and "proficiency" and "ability" are used interchangeably throughout the paper.

## II. IRT Basics

When students perform worse than expected on a test item, many physics educators would ask the same question: was it more about the question itself or more about their students? Unfortunately, there is no simple answer to this question. In classical test theory (CTT), question difficulty is defined based on the proportion of correct responses given by one particular group of examinees. In this definition, item parameters are inseparable from the group. For example, a low average score of an item may suggest that the item is too hard for the group, or it may also suggest that the group has a relatively low proficiency.

Item response theory handles this problem differently: the item difficulty and examinee proficiency are described with two separate parameters. By experimentally fixing one parameter, the other can be conditionally determined. A common IRT procedure is to produce a calibrated assessment instrument with a target population, which goes through the following steps:

1. Locate a population of large size. Assume the proficiency of the examinee population follows a standard normal distribution and choose a large sample of examinees from the population.

2. Assign a test to the chosen sample, and assume the sample's proficiency also follows a standard normal distribution. Determine the item parameters of each test item by fitting these parameters to produce the best agreement between the actual test performance of the sample and the predicted performance produced by the model.
3. The validated item parameters are then calibrated based on the target population and can be treated as independent of the particular samples of the population chosen for testing.

If we carefully choose the large sample in step 1, ensuring it is a representative sample of the target population, the calibration of item parameters based on this sample is often very close to the population's "true" parameters and can be used for assessment practices. Once calibrated, the parameters of the assessment instrument can be held constant and the instrument can be used to measure differences (in terms of proficiency levels) between different individuals and groups of students and between different performance states of the same students.

## A. The probabilistic framework of IRT

IRT models an examinee's behavior (score) with a probabilistic framework. In an item response model, an examinee's score on one item is a probability function depending on two sets of parameters: one set describing the examinee and one set describing the question. A general mathematic form of the IRT models can be written as

$$P_i(\theta) = P_i(Y_i = y_i | \{\theta\}, \{\delta_i\}) . \tag{1}$$

Here, $\{\theta\}$ is a set of parameters describing the examinee's characteristics, whereas $\{\delta_i\}$ is a set of parameters describing the features of the $i^{th}$ item. $Y_i$ is a variable representing the examinee's response to the $i^{th}$ item. If the examinee gives a specific response $y_i$ (i.e. $Y_i = y_i$), his/her score $P_i(\theta)$ is defined as the probability for him/her to make such a response given the characteristics of both the person and the item. Equation (1) gives a theoretical equation with no practical definition of the parameters. A function that follows the general form in Eq. (1) is called an *item characteristic function*.

The first developed concrete form of item response model is called the "normal ogive" model, which uses the normal ogive as its probability function[7]. In statistics textbooks, normal ogive is often called normal cumulative distribution function, which is the integral form of a normal distribution (see the top of Fig. 1 for an example of normal ogive). The model assumes that the proficiency or ability variable $\theta$ of the examinee population is normally distributed. A student's response to a multiple choice item is either correct or incorrect represented with $Yi=1$ or $Yi=0$, respectively. Following the general form of Eq. (1), the probability for an examinee with proficiency $\theta$ to respond correctly is

$$P_i(\theta) = P(Y_i = 1 | \theta) = \Phi\{-a_i(\theta - b_i)\}, \qquad (2)$$

where the parameters $a_i$ and $b_i$ correspond to the item parameter set $\{\delta_i\}$ in Eq. (1), and $\theta$ corresponds to $\{\theta\}$ in Eq. (1). Here $a_i$ is the discrimination parameter of the $i^{th}$ item, $b_i$ is the item difficulty, and $\Phi$ is the normal ogive. Item discrimination parameter $a_i$ describes the ability of an item to distinguish examinees with different proficiencies. Item difficulty parameter $b_i$ describes the difficulty of an item with a high value corresponding to a hard question. In the normal ogive model, the item discrimination parameter equals to the slope at the center of the normal ogive curve, and the difficulty of an item equals to the $\theta$ value at the center of the normal ogive curve. The top part of Fig. 1 gives an illustration of Eq. (2). Because the horizontal axis corresponds to ($\theta$-$b_i$), the normal ogive curve is centered at ($\theta$-$b_i$) = 0.

Equation (2) reflects the "competition" between a student's proficiency and an item's difficulty, i.e., the difference between proficiency and item difficulty ($\theta$-$b_i$) in conjunction with the item discrimination parameter $a_i$ determines the probability for the student to give a correct response to the $i^{th}$ question. In this manner, the features of the question and the examinee are separately defined and can be further manipulated. The interaction between the two set of parameters through the probability function determines the score.

Figure 1 illustrates the relation between the proficiency distribution and the possible outcome of the $i^{th}$ item under the model Eq. (2). To make the illustration simple, we choose three different subgroups of examinees, each with a normal proficiency distribution centered at $\theta_j$, $j$ = 1, 2, 3. The label 1, 2, and 3 corresponds to a low-proficiency, a medium-proficiency, and a high-proficiency group. The widths of all three distributions are set to be the same. The top part of Fig. 1 is a plot of the item characteristic function, Eq. (2), which centers at $\theta$-$b_i$ = 0. The centers of the normal distributions of the three groups are located at $\theta_1$-$b_i$ = -2, $\theta_2$-$b_i$ = 0, and $\theta_3$-$b_i$ = 2. We can see that the majority of group 3 has proficiency larger than the item difficulty (see the shaded areas under the normal distribution curves), while the opposite is true for group 1. In group 2 students' proficiencies are evenly distributed above and below the item difficulty. Since the probability for a student to have a specific outcome ($Y_i$=0 or 1) is a function of ($\theta$-$b_i$), we can expect that the probability for a student to give a correct answer to this question will be large, medium, and small for students in group 3, 2, and 1, respectively. In practice, an examinee population can be viewed as a collection of many such subgroups; the item characteristic curve plotted at the top portion of Fig. 1 gives the item response characteristics of a particular item tested with the entire population.

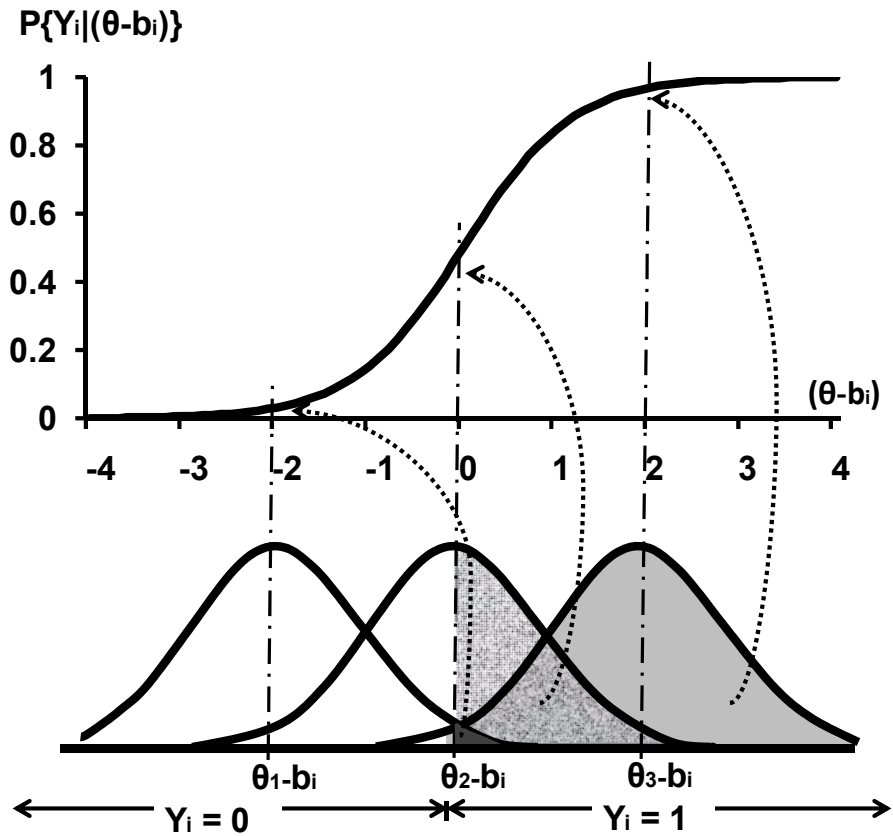

Fig. 1. Item characteristic function of the $i^{th}$ item and its relation with different examinee proficiency distributions. Top: item characteristic curve of the $i^{th}$ item. Bottom: Proficiency or ability distribution of three distinctive examinees. [1]

(1) A slight revision of Thissen & Orlando's Fig. 3.3, pp. 80 and Fig. 3.6, pp. 86, 2001.[7]

**B. The three-parameter item response model**

Historically, logistic functions are often used instead of the normal ogive for mathematical convenience. The difference between the logistic function and the normal ogive can be addressed by adding appropriate constants in the logistic function.[8] This substitution leads to a logistic family of item response models. One important member of this family, the three-parameter item response model, which can be applied to analyze multiple-choice test data, is the model we chose for our study.

In the three-parameter item response model, the probability of answering the $i^{th}$ item correctly for an examinee with proficiency $\theta$ is

$$P_i(\theta) = c_i + \frac{1-c_i}{1+\exp[-1.7a_i(\theta-b_i)]}, \qquad (3)$$

where $a_i$ and $b_i$ represents discrimination parameter and difficulty parameter of the $i^{th}$ item. The parameter $c_i$ is the lower asymptote of the item characteristic function as $\theta \to -\infty$, describing the probability of giving a correct answer by an examinee with extremely low proficiency. For a multiple-choice item, there is always a non-trivial probability to answer the question correctly through guessing, which is represented with $c_i$. The constant 1.7 is chosen to make the logistic function behave similarly to the normal ogive with the same set of parameters. The typical ranges of parameters $a_i$, $b_i$ and $c_i$ are determined empirically[11] and shown in Table I.

It is also assumed that the proficiency of an examinee is not changing from item to item; the probability of answering the $i^{th}$ item correctly for this examinee is in general related to his/her overall performance. Therefore, Eq. (3) should fit all the items of a test under a common $\theta$. Obviously, context features may play significant roles in changing student behaviors on questions designed based on the same scientific concept but with different contexts.[9] In this paper, we will not go into detail on the context dependence issue but rather apply the IRT models based on the original assumptions.

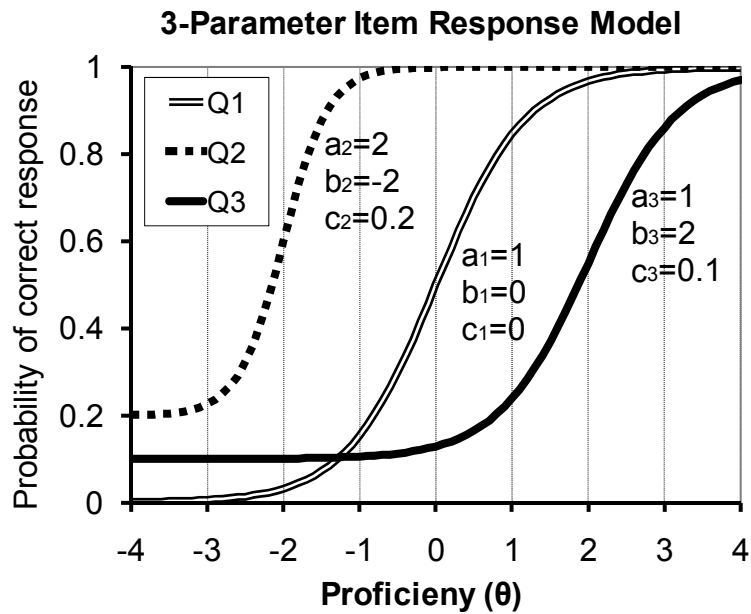

Fig. 2. Three-parameter item characteristic curves of three different items.

Figure 2 gives an example of item characteristic curves of three items with different parameters. The item difficulty parameter *b* determines the center location of each curve, the item discrimination parameter *a* determines the slope of the center part of each curve, and the guessing parameter c determines the lowest probability for a student to answer the question correctly. As shown by the curves, the most discriminative item is item 2, which rises steeply around the center distinguishing the examinee proficiency more clearly than the other two items would do. The items from the easiest to the hardest are item 2, item 1, and item 3. Their corresponding item difficulty parameters are $b_2 = -2$, $b_1 = 0$, and $b_3 = 2$. Item 1 has a guessing parameter value of 0, which suggests that the item may have a strong distracter and the chance of correct response for low proficiency students is fairly low.

### III. Data collection and analysis

At the Ohio State University, from September 2003 to June 2007, the students who enrolled in a calculus-based introductory mechanics course took the FCI in the second week and the last week of the quarter as pre- and post- diagnostics. The number of students tested is 2802 for pre- and 2729 for post- diagnostics. The data collection was a part of the regular lab activities until September 2007, when a reformed lab curriculum was implemented. The average pre- and post- FCI scores of each quarter (200~300 students per quarter) is fairly steady over time. We combined all the data over the years by pre- and post- tests.

In calibrating item parameters, the key point is to choose the appropriate sample to represent a general population. In the sample mentioned above, 2802 students participated in the FCI pre-testing with an average total score of 15.80 (out of 30) and a standard deviation of 6.05. The histogram of students' raw scores is fairly normal and the average score is about 50% of the maximum FCI score. These conditions make the pretest data ideal to fit the IRT model and calibrate FCI item parameters. In this paper, the three-parameter item response model is fitted to the FCI pre-test data.

Table I contains both the IRT item parameters estimated from data fitting and the traditional item parameters calculated using classical test theory (CTT).[10] The CTT parameters are listed here for reference. The Chi-square test statistics of overall fit of each item are listed as well. For IRT, the typical ranges of the parameters are decided empirically.[11] If a parameter's value exceeds the typical range, further investigation is often suggested. For CTT, the possible values for the item discrimination and item difficulty are both in the range of [0, 1], and computed values based on the pre-test FCI data are also listed in the table. For Chi-square statistics, the typical range suggests a good model-data fit. Most of the IRT parameters of FCI fall into the typical range, except that three questions (5, 13, and 18) have discrimination parameters greater than 2, and that the guessing chance of question 16 is greater than 30%.

Table I. Fitting 3-parameter item response model to FCI pre-test data (N=2802)

| Item parameters | Discrimination | | Difficulty | | Guessing [3] | Overall Fit |
|---|---|---|---|---|---|---|
| | IRT - $a$ | CTT - disc.[1] | IRT - $b$ | CTT - diff.[2] | IRT - $c$ | Chi$^2$ statistic |
| Typical range | [0, 2] | [0, 1] | [-3, 3] | [0, 1] | [0, 30%] | [0, 1] |
| F1 | 0.72 | 0.36 | -1.53 | 0.85 | 21% | 0.12 |
| F2 | 0.81 | 0.45 | 0.57 | 0.47 | 17% | 0.46 |
| F3 | 0.55 | 0.38 | -0.94 | 0.72 | 13% | 0.20 |
| F4 | 1.19 | 0.49 | 0.80 | 0.38 | 15% | 0.44 |
| F5 | 2.16 | 0.51 | 1.08 | 0.24 | 8% | 0.89 |
| F6 | 0.59 | 0.31 | -1.70 | 0.85 | 23% | 0.08 |
| F7 | 0.59 | 0.36 | -0.81 | 0.74 | 24% | 0.12 |
| F8 | 0.71 | 0.43 | -0.19 | 0.64 | 20% | 0.23 |
| F9 | 0.81 | 0.39 | 0.88 | 0.48 | 26% | 0.25 |
| F10 | 0.93 | 0.48 | -0.72 | 0.72 | 9% | 0.47 |
| F11 | 1.54 | 0.57 | 0.75 | 0.32 | 7% | 0.67 |
| F12 | 0.66 | 0.37 | -1.32 | 0.80 | 13% | 0.20 |
| F13 | 2.83 | 0.64 | 0.58 | 0.38 | 11% | 0.95 |
| F14 | 0.66 | 0.45 | -0.47 | 0.63 | 8% | 0.41 |
| F15 | 0.63 | 0.36 | 1.29 | 0.34 | 13% | 0.57 |
| F16 | 0.87 | 0.42 | -0.17 | 0.70 | 34% | 0.14 |
| F17 | 1.44 | 0.44 | 1.39 | 0.18 | 6% | 0.81 |
| F18 | 2.72 | 0.58 | 0.87 | 0.28 | 9% | 0.69 |
| F19 | 0.53 | 0.38 | -0.64 | 0.67 | 13% | 0.26 |
| F20 | 0.65 | 0.44 | -0.49 | 0.64 | 9% | 0.42 |
| F21 | 0.63 | 0.37 | 0.75 | 0.49 | 23% | 0.29 |
| F22 | 0.89 | 0.45 | 0.42 | 0.55 | 26% | 0.34 |
| F23 | 0.75 | 0.49 | -0.08 | 0.54 | 5% | 0.56 |
| F24 | 0.94 | 0.45 | -0.94 | 0.77 | 13% | 0.34 |

| F25 | 1.95 | *0.47* | 1.19  | *0.23* | 10% | 0.72 |
| F26 | 1.74 | *0.52* | 1.33  | *0.14* | 2%  | 0.90 |
| F27 | 0.53 | *0.35* | -0.64 | *0.71* | 23% | 0.15 |
| F28 | 1.23 | *0.57* | 0.46  | *0.44* | 13% | 0.48 |
| F29 | 0.32 | *0.29* | -0.15 | *0.57* | 11% | 0.33 |
| F30 | 1.67 | *0.54* | 0.87  | *0.30* | 9%  | 0.67 |

(1) In classical test theory, the item discrimination is the point-biserial correlation coefficient between the item scores (0 or 1) and the total test scores.[10]

(2) In classical test theory, the item difficulty is the percentage of correct response, so a low value indicates a high item difficulty.

(3) In classical test theory, there is no formal definition of guessing chance. Guessing chance of a multiple-choice question is often estimated as (1/ number of choices), which is 20% for every item in FCI. This estimation implies the probability of choosing each choice is equal.

## IV. Model validation

### A. Unidimensionality and item independence

Educational tests are usually given to reveal the examinee differences on the abilities or skills measured by test items. If a test is intended to measure the proficiency level of one skill, it is said to be *unidimensional*. Unidimensionality is one basic assumption in the 3-parameter item response model. This one-dimensional structure is a hypothetical cognitive construct. Practically, students' knowledge always involves multiple substructures; however, it is possible to develop a test that the measured skills are well connected forming a single general skill dimension. Therefore, with a given test intended for IRT analysis, the first step is to evaluate unidimensionality of the test data.

Local item independence is the other important assumption for all item response models. Local item independence assumes that for examinees at a fixed proficiency level $\theta_{fix}$, their performance on one item is not conditionally dependent on their performance on another item. In one dimensional proficiency space, local independence assumption is equivalent to the unidimensionality assumption. If a test is unidimensional, then the proficiency variable $\theta$ should be the only variable that accounts for the variance caused by differences in students' innate abilities. If, for students with the same proficiency $\theta_{fix}$, the score of one item is dependent on the score of another item of the test, there must be additional cognitively based variables describing such mechanisms, which violates the unidimensionality assumption.[10, 12] Therefore, if the unidimensionality assumption is satisfied, the local item independence assumption is automatically verified.

The dimensionality of a test can be studied by analyzing the correlations between results of different items, which are often put into a correlation matrix. An eigenvalue analysis of the correlation matrix can show how much a set of eigenvectors (or factors) can represent the variances in the test data. The set of eigenvectors are built from the original items, however, they do not represent the original items directly. If a dominant eigenvalue/eigenvector (one that is an order of magnitude larger than others) is found, it indicates that most of the variance of the correlation matrix can be described with that eigenvector and dimensions of the correlation matrix is significantly reduced. A proving case of a unidimensional correlation matrix should have one eigenvalue orders of magnitude larger than the rest. A useful tool called "scree plot", which shows the sorted eigenvalues from large to small, is often used to qualitatively represent and analyze the dimensionality of a correlation matrix.[13]

We obtained the correlation matrix for the pre-test data (N=2802). In the computation, we treat each item as an independent dimension resulting in one 30x30 correlation matrix. Since students' responses to an item are in a binary form (1 or 0), we used the tetrachoric correlation to compute the correlation matrix.[14, 15]

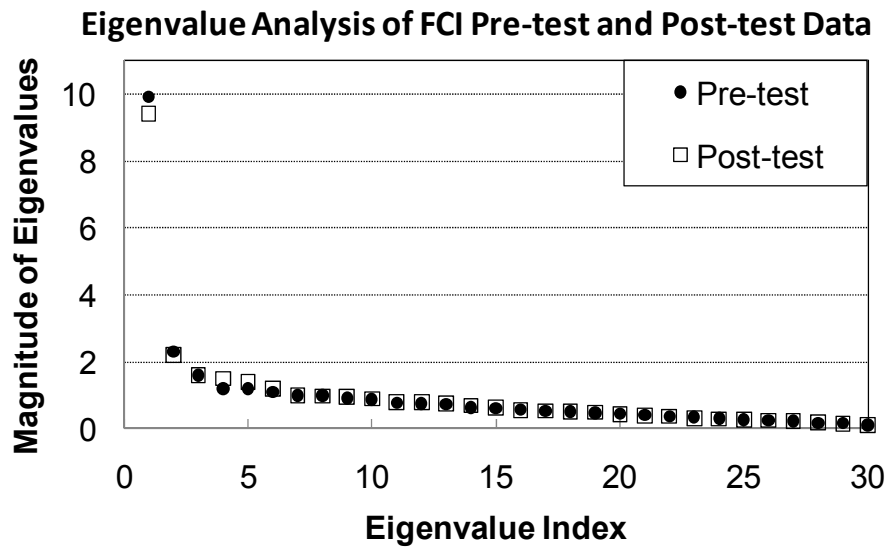

Figure 3. Eigenvalue analysis of the correlation matrices of FCI pre-test (N=2802) and post-test (N=2729) data.

To perform IRT analysis of FCI data, the unidimensionality assumption should be verified for both the pre-test and the post-test data. For this purpose the correlation matrix for the post-test data (N=2729) is calculated using the same method. The scree plot of eigenvalues of both correlation matrices are shown in Fig. 3. The horizontal axis indexes the eigenvalues from large to small, which are not connected to the original item numbers. We notice that the eigenvalues of pre- and post-test data are similar for all 30 ordered eigenvalues. In both pre and post testing, the first eigenvalue is significantly (about 5 times) larger than the rest, which suggests a single proficiency variable accounting for a significant portion of the variances. The similarities between the eigenvalues also suggests that there are no obvious changes from pre- to post-testing in students' cognitive constructs responsible for the variances in the data. The results indicate that it is reasonable to assume unidimensionality for IRT analysis of FCI data.

## B. Goodness of IRT fit

A Pearson chi-square statistic is often used to evaluate the goodness of fit between IRT model and data (see Eq. (4)).[16]

$$Q_i = \frac{1}{(J-3)} \sum_{j=1}^{J} \frac{N_j (O_{ij} - E_{ij})^2}{E_{ij}(1 - E_{ij})}. \tag{4}$$

The computation is done with the following procedure. We divide the range of proficiency scale into J bins, with each bin including approximately the same number of examinees. For each item $i$ we compare the observed proportions of correct response $O_{ij}$ with the expected proportions of correct response $E_{ij}$ in bin $j$, where $N_j$ is the number of examinees in bin $j$. If the model fits the data well, each $Q_i$ follows a $\chi^2(1)$ distribution and gives the goodness of fit of an item.[16] Since the parameters of the three-parameter item response model are obtained by fitting the model to the data of the whole test (all 30 questions), the value of $Q_i$ for each item may not be the smallest that it can get, but the sum of all $Q_i$'s is approaching minimum through the fit. The chi-square statistics for all 30 questions are listed in Table I. Here all the Q values are less than 1, falling within the 68% quantile of the $\chi^2(1)$ distribution, which is the commonly accepted critical value in a one-degree chi-square test. None of the Q values is unusual assuming the model correct, hence gives us no reason to seriously doubt the model.[17] We then believe none of the 30 questions is significantly misfitted. Therefore, the goodness of fit analysis suggests that it is acceptable to fit the three-parameter item response model with pre FCI data.

## V. FCI measurement metrics

Based on large scale data of a representative population, we obtained good estimations of item characteristic parameters for each of the FCI questions with the three-parameter item response model. These parameters and the model can be used in practice for a variety of education and research purposes. For easy applications, we plotted the fitted item characteristic function (ICF) of each of the questions together to form a measurement metric (the FCI-metric) that allows users to conveniently compare and reference the different features of the questions and the differences in performances of different students.

The FCI-metric is shown in Fig. 4. The dark thick sigmoid curve in each small plot shows the item characteristic function (ICF) of a question, with item parameters a, b, and c labeled on the graph. The y-axis is the probability of giving a correct response, and the x-axis is the proficiency θ. The discrete dots on each graph represent the observed probability of correct answers from the students at different proficiency level. We binned all 2802 data points into 56 proficiency bins, each containing approximately 50 data points, therefore each plot has 56 short grey lines (error bars). The error bar represents the standard deviation of the percentage of the correct response within each bin.

As illustrated in Fig. 2, the ICF shifts to the left if the item is easy, and to the right if the item is difficult. Results in Fig. 4 show that questions 1 and 6 are the easiest, while questions 25 and 26 are the hardest. The slope of the curve represents the item discrimination, i.e. the ability of an item to distinguish between high proficiency students and low proficiency students. On this aspect, questions 5, 13 and 18 are better than others. The ICF of question 29 is almost flat, suggesting its low ability in discrimination. The left asymptote of the ICF curve represents the guessing chance (in Fig.4, the x-axis ranges from -3 to +3, hence the intercept doesn't always show the left asymptote). The results show that on questions 23 and 26, the low-proficiency students have no more than 5% chance to guess correctly. On the other hand, in answering question 16, almost 1/3 of the low-proficiency students can make a correct guess.

To study how students' total scores on the FCI test are related to $\theta$, we analyzed the relation between the individual students' total raw scores (percentage correct) and their proficiency $\theta$'s obtained through IRT fits. The results show that FCI raw score $S$ and the proficiency $\theta$ have an approximately linear relationship with the correlation coefficient $R^2$ equal to 0.994, as shown in Fig. 5. The fitted linear relation is given below:

$$\widetilde{\theta} = 0.147 * S - 2.324 . \tag{5}$$

The error bar in Fig. 5 is the standard deviation of the proficiency $\theta$ within the students of the same raw score. Students with the same raw score don't have to have the same proficiency depending on the specific sets of questions they got right and wrong.

It is worth noticing that although on individual questions the fitted item characteristic functions shown clear nonlinear shapes, the total FCI score corresponds linearly to the proficiency obtained through fitting the individual questions. This result suggests that using the total FCI score to evaluate students is consistent with the IRT analysis outcome.

Eq. (5) can be used to quickly estimate students' proficiency from their total FCI raw scores. Given the estimated proficiency, teachers can then use the FCI-metric (Fig. 4) to compare their students' performances with the large scale norm on each of the FCI questions. The results can provide formative guidance to the instructors and can also be used to evaluate the impact of education interventions.

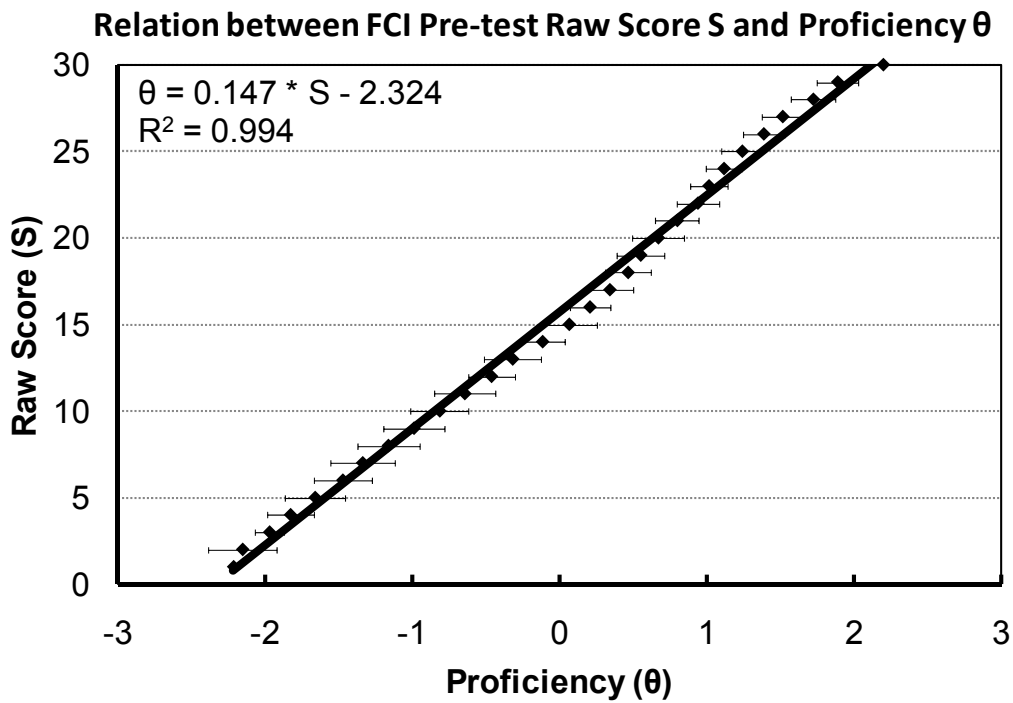

Figure 5. Relation between FCI pre-test raw score and proficiency $\theta$ (N=2802).

## VI. Summary

This paper presents a complete IRT analysis of FCI data. A three-parameter item response model is used to analyze the FCI pre-test data. The results are compiled to form a FCI-metric, which provides a clear and convenient way to evaluate the question features and to compare students' test data.

In using the FCI-metric as a reference in teaching practice, the probabilistic nature of the underlying IRT model should always be kept in mind by users. The FCI-metric gives a general summarization of students' behavior, which can involve significant uncertainties when used to predict single student's performance on a test. The prediction will get much improved with a small group of students. When the group gets large, teachers may want to break it down into smaller groups based on their average scores.

We believe that the presented analysis for building model based measurement metrics can be useful for physics education researchers who design tests and want to disseminate their tests for general applications. In the dissemination process of a research-based test, one question often encountered is: How do we interpret the test score? The educators want to know what is the range of an acceptable student score, which questions are harder than others, how do the questions discriminate students, and based on their own students' entry level, what could be called a significant change in pre-/post- diagnostics[18]. The technique presented in this paper can be applied to a sample as small as 200-300 students; however, it is recommended to use a much larger sample that is also representative of the target population. By applying the technique, the item parameters of a developed test can be calibrated based on a large scale sample of a target population. Therefore, other educators and researchers with different samples of a similar population can use these parameters to study their students.

**Acknowledgments**

The authors would like to thank all members of the Physics Education Research Group at The Ohio State University for their long-term support and valuable suggestions on this project. The research is supported in part by NIH Award RC1RR028402 and NSF Award DUE-0633473. Any opinions, findings, and conclusions or recommendations expressed in this paper are those of the author(s) and do not necessarily reflect the views of the National Institute of Health and National Science Foundation.

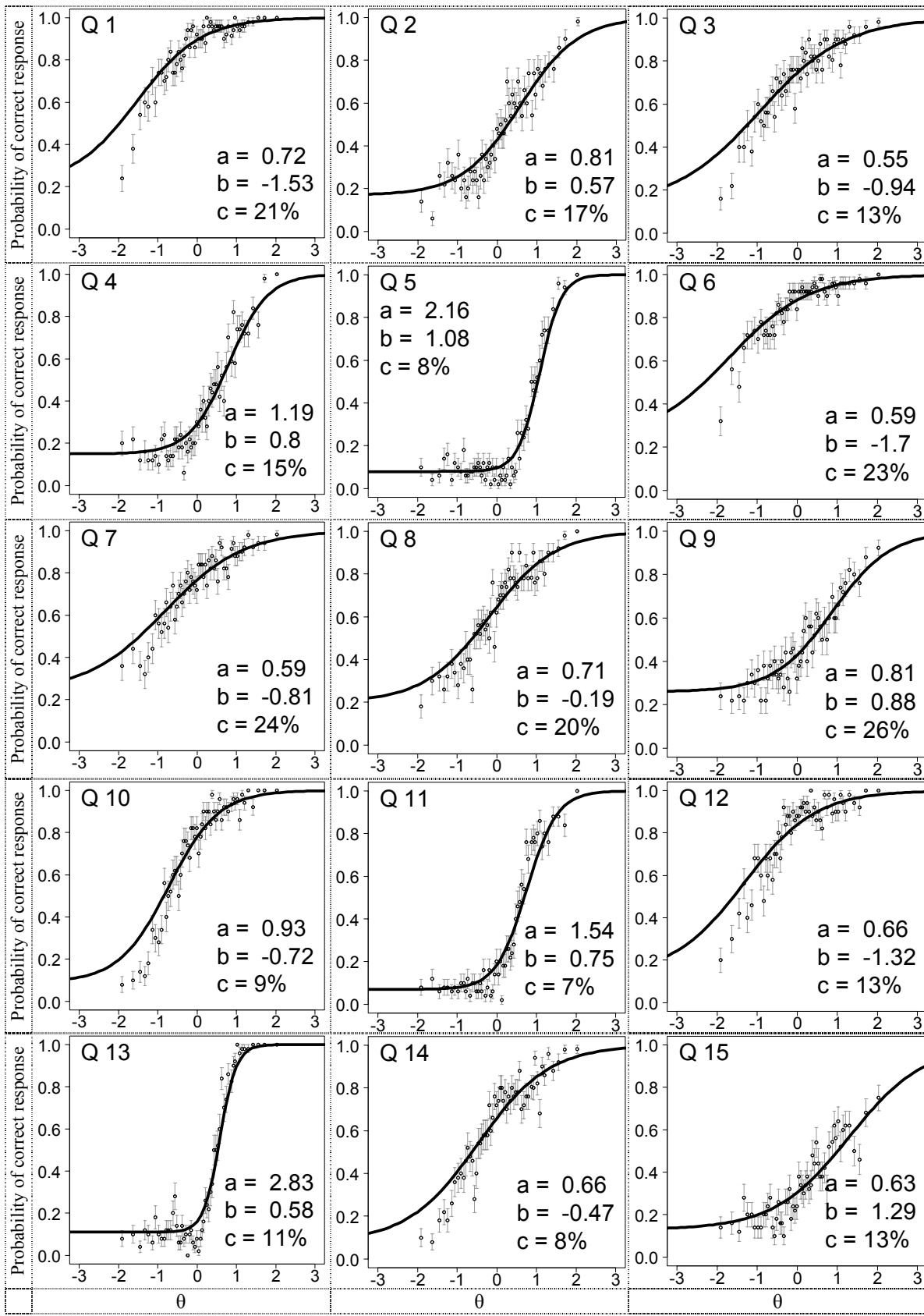

Figure 4.(a) FCI-metric: graphical representation of ICFs of FCI questions 1-15 (N=2802)

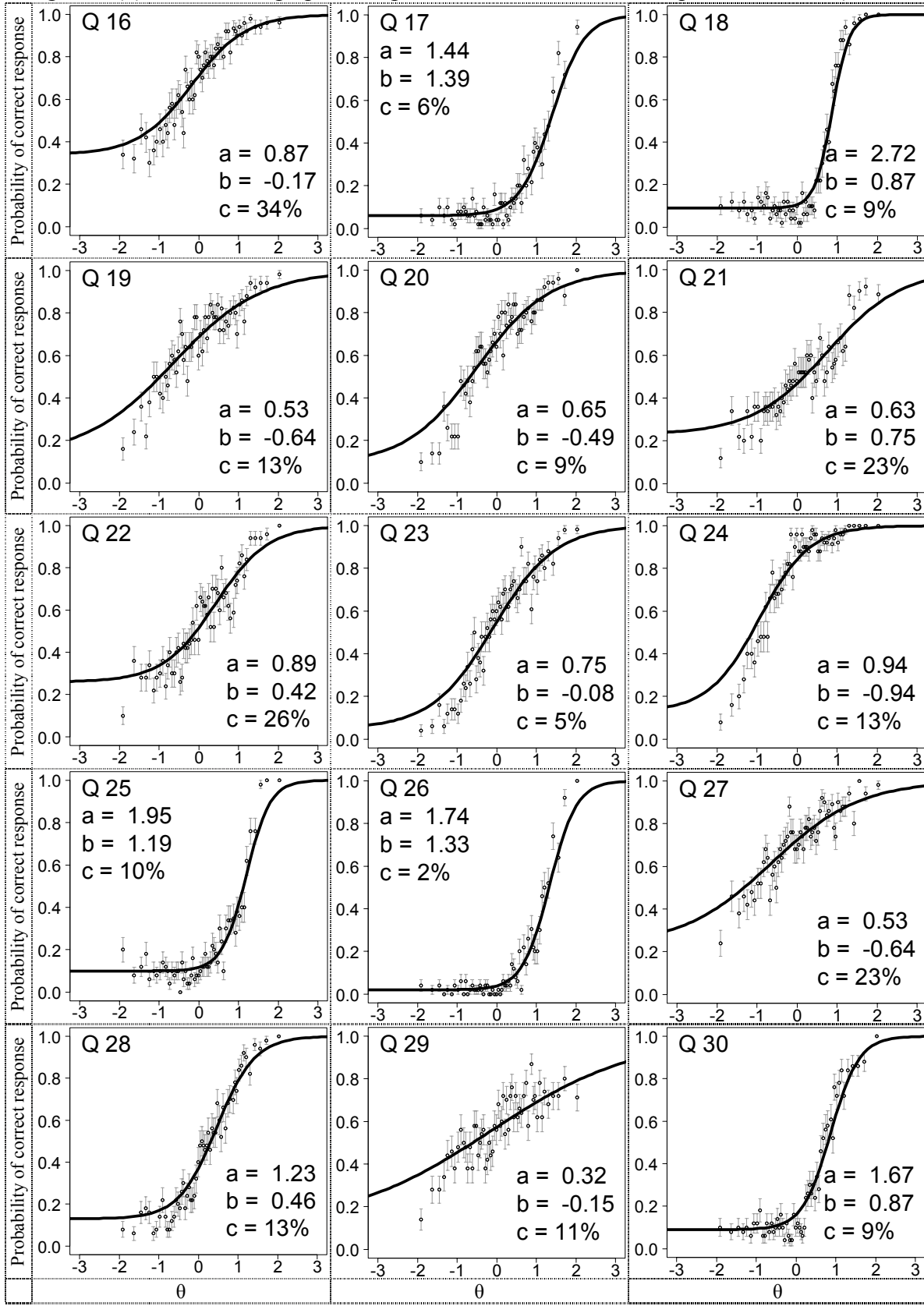

Figure 4.(b) FCI-metric: graphical representation of ICFs of FCI questions 16-30 (N=2802)

**References and Endnotes**

[14] Drasgow, F. (1988). Polychoric and polyserial correlations. In Kotz, L, and Johnson, NL (Eds.), Encyclopedia of statistical sciences. Vol. 7 (pp. 69-74). New York: Wiley.

[15] Harris, B. (1988). Tetrachoric correlation coefficient. In Kotz, L, and Johnson, NL (Eds.), Encyclopedia of statistical sciences. Vol. 7 (pp. 69-74). New York: Wiley.

[16] Yen, W. (1981). Using simulation results to choose a latent trait model. Applied Psychological Measurement, v5, 245-262.

[17] Rice, J. (1995). Mathematical statistics and data analysis. 2$^{nd}$ Ed. Belmont, CA: Duxbury Press.

[18] An example can be found in this paper: Lee, Y., Palazzo, D., Warnakulasooriya, R. and Pritchard, D. (2008). Measuring student learning with item response theory. Physics Review Special Topics – Physics Education Research, v4, 010102.
- 18 -